\documentclass[10pt,journal,compsoc]{IEEEtran}

\IEEEoverridecommandlockouts

\usepackage{cite}
\usepackage{booktabs} 
\usepackage{graphicx}
\usepackage{array}
\usepackage{url}
\usepackage{fancybox}
\usepackage{multirow}
\usepackage{amsmath,amssymb,amsfonts}
\usepackage{color}
\usepackage[table,xcdraw]{xcolor}
\usepackage{fancyhdr}
\usepackage{subfig}
\usepackage{diagbox}
\usepackage{bbding}
\usepackage{placeins}
\usepackage{balance}
\usepackage{hyperref} 
\usepackage{enumitem}
\usepackage[linesnumbered, ruled,vlined]{algorithm2e}
\def\BibTeX{{\rm B\kern-.05em{\sc i\kern-.025em b}\kern-.08em
		T\kern-.1667em\lower.7ex\hbox{E}\kern-.125emX}}

\usepackage{algorithm2e}
\usepackage{multirow}
\usepackage{algpseudocode}
\usepackage{tikz}
\usepackage{amsmath}
\usepackage{makecell}
\usepackage{diagbox}
\usepackage{url}
\newcommand{\circnumber}[1]{\lower.75ex\hbox{\tikz\draw (0pt, 0pt)%
    circle (.47em) node {\makebox[.15em][c]{\small #1}};}}
\newcommand{\ballnumber}[1]{\lower.75ex\hbox{\tikz\fill(0pt, 0pt)%
    circle (.5em) node {\makebox[.15em][c]{\small \textcolor{white}{#1}}};}}
\newcommand{\dcircnumber}[1]{\lower.75ex\hbox{\tikz\draw(0pt, 0pt)%
    circle (.47em) circle (.37em) node {\makebox[.15em][c]{\small #1}};}}

\newcommand{\emptycirc}[1]{\lower.75ex\hbox{\tikz\draw (0pt, 0pt)%
    circle (.47em) node {\makebox[.15em][c]{\small \textcolor{white}{#1}}};}}
\newcommand{\emptyball}[1]{\lower.75ex\hbox{\tikz\fill(0pt, 0pt)%
    circle (.5em) node {\makebox[.15em][c]{\small #1}};}}
\newcommand{\emptydcirc}[1]{\lower.75ex\hbox{\tikz\draw(0pt, 0pt)%
    circle (.47em) circle (.37em) node {\makebox[.15em][c]{\small \textcolor{white}{#1}}};}}
            
\usepackage{listings, xcolor}
\definecolor{verylightgray}{rgb}{.97,.97,.97}
\lstdefinelanguage{Solidity}{
  keywords=[1]{anonymous, assembly, assert, balance, break, call, callcode, case, catch, class, constant, continue, constructor, contract, debugger, default, delegatecall, delete, do, else, emit, event, experimental, export, external, false, finally, for, function, gas, if, implements, import, in, indexed, instanceof, interface, internal, is, length, library, log0, log1, log2, log3, log4, memory, modifier, new, payable, pragma, private, protected, public, pure, push, require, return, returns, revert, selfdestruct, send, solidity, storage, struct, suicide, super, switch, then, this, throw, transfer, true, try, typeof, using, value, view, while, with, addmod, ecrecover, keccak256, mulmod, ripemd160, sha256, sha3}, 
  keywordstyle=[1]\color{blue}\bfseries,
  keywords=[2]{address, bool, byte, bytes, bytes1, bytes2, bytes3, bytes4, bytes5, bytes6, bytes7, bytes8, bytes9, bytes10, bytes11, bytes12, bytes13, bytes14, bytes15, bytes16, bytes17, bytes18, bytes19, bytes20, bytes21, bytes22, bytes23, bytes24, bytes25, bytes26, bytes27, bytes28, bytes29, bytes30, bytes31, bytes32, enum, int, int8, int16, int24, int32, int40, int48, int56, int64, int72, int80, int88, int96, int104, int112, int120, int128, int136, int144, int152, int160, int168, int176, int184, int192, int200, int208, int216, int224, int232, int240, int248, int256, mapping, string, uint, uint8, uint16, uint24, uint32, uint40, uint48, uint56, uint64, uint72, uint80, uint88, uint96, uint104, uint112, uint120, uint128, uint136, uint144, uint152, uint160, uint168, uint176, uint184, uint192, uint200, uint208, uint216, uint224, uint232, uint240, uint248, uint256, var, void, ether, finney, szabo, wei, days, hours, minutes, seconds, weeks, years},  
  keywordstyle=[2]\color{teal}\bfseries,
  keywords=[3]{block, blockhash, coinbase, difficulty, gaslimit, number, timestamp, msg, data, gas, sender, sig, value, now, tx, gasprice, origin},  
  keywordstyle=[3]\color{violet}\bfseries,
  identifierstyle=\color{black},
  sensitive=false,
  comment=[l]{//},
  morecomment=[s]{/*}{*/},
  commentstyle=\color{red}\ttfamily,
  stringstyle=\color{red}\ttfamily,
  morestring=[b]',
  morestring=[b]"
}
\begin{document}
	
\title{SmartOracle: Generating Smart Contract Oracle via Fine-Grained Invariant Detection}
\author{Jianzhong Su, Jiachi Chen, Zhiyuan Fang, Xingwei Lin, Yutian Tang and Zibin Zheng,~\IEEEmembership{Fellow,~IEEE,}
\IEEEcompsocitemizethanks{\IEEEcompsocthanksitem Jianzhong Su, Jiachi Chen, Zhiyuan Fang and Zibin Zheng Ye are with School of Software Engineering, Sun Yat-sen University, China. \protect\\
E-mail: \{sujzh3, fangzhy27\}@mail2.sysu.edu.cn
E-mail: \{chenjch86, zhzibin\}@mail.sysu.edu.cn

\IEEEcompsocthanksitem Xingwei Lin is with the College of Computer Science and Technology, Zhejiang University, China.\protect\\
E-mail: xwlin.roy@gmail.com
\IEEEcompsocthanksitem Yutain Tang is with the School of Computing Science, University of Glasgow, United Kingdom.\protect\\
E-mail: yutian.tang@glasgow.ac.uk
			
\IEEEcompsocthanksitem Zibin Zheng is the corresponding author.}
  }
	
	\markboth{}%
	{Shell \MakeLowercase{\textit{et al.}}: Bare Demo of IEEEtran.cls for Computer Society Journals}

	\IEEEtitleabstractindextext{%
            \begin{abstract}

As decentralized applications (DApps) proliferate, the increased complexity and usage of smart contracts have heightened their susceptibility to security incidents and financial losses. Although various vulnerability detection tools have been developed to mitigate these issues, they often suffer poor performance in detecting vulnerabilities, as they either rely on simplistic and general-purpose oracles that may be inadequate for vulnerability detection, or require user-specified oracles, which are labor-intensive to create. In this paper, we introduce SmartOracle, a dynamic invariant detector that automatically generates fine-grained invariants as application-specific oracles for vulnerability detection. From historical transactions, SmartOracle uses pattern-based detection and advanced inference to construct comprehensive properties, and mines multi-layer \textit{likely} invariants to accommodate the complicated contract functionalities. After that, SmartOracle identifies smart contract vulnerabilities by hunting the violated invariants in new transactions.
In the field of invariant detection, SmartOracle detects 50\% more ERC20 invariants than existing dynamic invariant detection and achieves 96\% precision rate. Furthermore, we build a dataset that contains vulnerable contracts from real-world security incidents. SmartOracle successfully detects 466 abnormal transactions with an acceptable precision rate 96\%, involving 31 vulnerable contracts. The experimental results demonstrate its effectiveness in detecting smart contract vulnerabilities, especially those related to complicated contract functionalities.  

\end{abstract}
	
        }

	
	\maketitle
	\IEEEdisplaynontitleabstractindextext


	\section{Introduction} \label{sec:intro}


Smart contracts are computer programs that run on a blockchain network. They act as the rules for DApps, which manage their funds and status without any central authority~\cite{zheng2020overview}. However, smart contracts are susceptible to security incidents due to various types of vulnerabilities, such as \textit{Reentrancy}~\cite{luu2016making} and \textit{Price Manipulation}~\cite{wu2021defiranger}. 
According to Slowmist~\cite{slowmisthacked}, these attacks have resulted in a total loss of approximately 32 billion USD on blockchain platforms, highlighting significant security challenges within the ecosystem.

To detect these vulnerabilities, various tools have been developed by employing a variety of technologies, such as static analysis~\cite{tsankov2018securify}, fuzzing~\cite{su2022effectively}, and formal verification~\cite{wang2023automated}. 
Despite these efforts, the detection of smart contract vulnerabilities, particularly those with complex functionalities, remains ineffective in real-world scenarios~\cite{zhang2023demystifying}. This ineffectiveness arises primarily from two reasons. 
\textit{First}, many existing tools~\cite{su2021evil, choi2021smartian, bose2022sailfish, nguyen2020sfuzz, zhang2020txspector, feist2019slither, jiang2018contractfuzzer} adopt simplistic and general-purpose oracles to cover all patterns of complex vulnerabilities. However, these tools may fail to reveal many vulnerabilities, as analyzing them usually requires application-specific oracles. For example, some tools use ``check-effect-interaction" oracle for \textit{Reentrancy} detection~\cite{luu2016making}; however, the complicated contract functionalities render such a simple oracle ineffective, resulting in a high rate of false positives~\cite{zheng2023turn}.
\textit{Second}, although some tools~\cite{duan2022towards, grieco2020echidna, wang2020formal} allow users to design custom oracles to detect vulnerabilities, these oracles require expert knowledge and significant human effects based on users' understanding of the contract. Consequently, the current reliance on inadequately tailored oracles limits the effectiveness of existing tools in identifying complex vulnerabilities within smart contracts.

\noindent\textbf{Our solution.} In this paper, we aim to propose a method to automatically generate application-specific oracles for detecting complex vulnerabilities within contracts. \textit{Program invariants} are properties that remain constant at certain execution points and naturally act as rules and oracles for smart contracts. For example, a smart contract is expected to hold the invariant that ``the tokens deposited into the contract are equal to the total amount claimed by users'' during normal execution. However, this invariant is violated in contracts suffering from a ``fake deposit'' vulnerability~\cite{thorchainhacked}, where the amount of tokens deposited is significantly lower than the attacker claims. Given the numerous transactions stored on the blockchain, we perform dynamic invariant detection on transactions to obtain \textit{likely} invariants~\cite{ernst2007daikon} of smart contracts and serve them as oracles for detecting vulnerabilities. Furthermore, complicated contract functionalities always require fine-grained invariants to fit. Specifically, some functions implement different functionalities with multiple branches, so we need to capture the corresponding fine-grained invariants for each branch. (detailed in Section~\ref{sec:motivation_insight}). 

We introduce SmartOracle, a tool designed to detect complex vulnerabilities in smart contracts via fine-grained dynamic invariant detection. SmartOracle starts the detection by analyzing historical transactions of a target contract and recording their execution traces (e.g., the functions, state variables, and executed branches). Based on the execution traces, SmartOracle delineates invariants of three distinct layers: \textit{Contract}, \textit{Function}, and \textit{Branch}, for accurately fitting contract behavior. At each layer, SmartOracle employs pattern-based property detection and advanced property inference to construct properties.
Considering that some historical transactions may be initiated by abnormal behaviors, which could introduce noise for the calculation of invariants. For instance, the contract had been attacked due to the abnormal state maintained by the manager but is still active after recovering to the normal state~\cite{qubithacked}. Thus, SmartOracle reserves properties that satisfy $threshold$ of historical transactions as \textit{likely} invariants to enhance robustness, thereby minimizing the impact of anomalies hidden in historical transactions. Based on the mined invariants, SmartOracle performs run-time verification on new transactions and reports their violated cases, indicating abnormal behavior or vulnerabilities triggered in contracts.

To evaluate SmartOracle, we first assess its ability to mine invariants. Utilizing a benchmark of 246 ERC20 contracts~\cite{liu2022invcon}, SmartOracle identifies 50\% more ERC20 invariants than the existing dynamic invariant detector, achieving a precision rate of 96\%. 
Furthermore, we create a dataset of 65 vulnerable contracts derived from real-world security incidents, most of which contain vulnerabilities with complicated functionalities. We employ the detected invariants as oracles to reveal smart contract vulnerabilities and the abnormal transactions they cause. SmartOracle successfully reveals 31 vulnerabilities, involving 466 transactions that violate the mined invariants. In addition, the efficiency of SmartOracle satisfy the requirement of real-time detection on Ethereum. These experimental results demonstrate the effectiveness of SmartOracle in vulnerability detection and run-time verification.

In summary, our main contributions are as follows: 
\begin{itemize}
    \item We propose SmartOracle, a fine-grained dynamic invariant detector for smart contracts. It mines fine-grained invariants and serves them as application-specific oracles for detecting complex vulnerabilities.
    \item We evaluate the ability of invariant detection, SmartOracle detects 50\% more ERC20 invariants than existing dynamic invariant detector and achieves 96\% precision.
    \item We construct a benchmark containing 65 real-world vulnerable contracts, most of which are complex vulnerabilities. SmartOracle reveals 31 vulnerabilities and 466 related abnormal transactions. 
    \item We publish the code of SmartOracle and related datasets in \url{https://github.com/Demonhero0/SmartOracle}.
    
\end{itemize}

	\section{Background} \label{sec:bkg}

\subsection{Smart Contracts} \label{sec:bkg_sc}

Smart contracts are Turing-complete programs running on top of the blockchain. They are typically written in a high-level programming language, such as Solidity~\cite{solidity}. Through compilation, we can obtain their Application Binary Interface (ABI) and bytecode~\cite{ethereum}. 
After deployment, smart contracts run within Ethereum Virtual Machine (EVM) and maintain their account states, which store Ether balances and \textit{storage} with a data structure serialized as Recursive Length Prefix (RLP)~\cite{ethereum}. In particular, \textit{storage} is a mapping between \textit{bytes32} keys to storage slots that record the state variables in \textit{bytes32}. 


\subsection{Transactions} \label{sec:bkg_tx}

Transactions are responsible for transferring Ethers or invoking smart contracts. They include information such as sender, receiver, amount of Ethers, and data recording the invoked function and its parameters. As smart contracts can initiate transactions, a transaction can derive multiple transactions arranged in a tree structure. For simplicity, transactions can be divided into two types, i.e., external transactions (initiated by external actors) and internal transactions (initiated by contracts). During transaction execution, smart contracts can emit predefined events to record specific behaviors, which contain address (i.e., the emitter), topics (i.e., the signature and indexed parameters) and data (i.e., non-indexed parameters). 


\begin{figure}
\setlength{\abovecaptionskip}{0.05cm}
\begin{lstlisting}[language=Solidity,mathescape]
contract ERC20Token {
// Contract Inv#1: totalSupply = SUM(balances)
  uint totalSupply;
  mapping(address => uint) balances;
  function transfer(address to, uint amt) {
// Function Inv#2: Pre(SUM(balances)) = Post(SUM(balances))
    if (balances[msg.sender] >= amt) {
// Branch Inv#3: amt=Post(balances[to]) - Pre(balances[to])
        balances[msg.sender] -= amt;
        balances[to] += amt;
    }
  }
}
\end{lstlisting}
\caption{The Invariants in ERC20 Contract.}
\label{fig:erc20}
\end{figure}

\subsection{Invariants} \label{sec:bkg_inv}

For a program, an invariant is a property (i.e., a logical formula) that remains true at certain points during execution. This property is important to the program and is widely used in bug detection, etc~\cite{ernst2007daikon}. In this paper, we focus on three layers of smart contract invariants, we illustrate them by Fig.~\ref{fig:erc20}:

\begin{itemize}
\item \textit{Contract Invariant}: the property holds for any call to the contract. The ERC20 contract uses \textit{totalSupply} to record the sum of users' token balance, it satisfies the property \textit{Contract Inv\#1}.
\item \textit{Function Invariant}: the property holds for any call to the function. The function \texttt{\small transfer} is responsible for the transfer of the token between users, which would not change the sum of the token balance of users. It satisfies the property \textit{Function Inv\#2}.
\item \textit{Branch Invariant}: the property holds for the execution branch. The function \texttt{\small transfer} only changes the token balance of the users when executing the branch with condition $balances[msg.sender] \geq amt$ (line 7). So \textit{Branch Inv\#3} only holds in the specific branch. 

\end{itemize}


	\section{Motivation} \label{sec:motivation}

In this section, we use a motivating example to illustrate the challenges of detecting vulnerabilities related to complicated contract functionality. Then, we introduce how to use invariant to address these challenges.


\begin{figure}
\setlength{\abovecaptionskip}{0.05cm}
\begin{lstlisting}[language=Solidity,mathescape]
contract Example {
  function deposit(address vault address ast, uint amt) {
    uint safeAmt;
    if (ast == RUNE) { // Branch#1
      safeAmt = amt;
      RUNE.transferFrom(msg.sender, vault, amt);
// Inv#1:amt=Post(token[ast][vault])-Pre(token[ast][vault])
    } else { // Branch#2
      safeAmt = amt;
      ast.transferFrom(msg.sender, this, amt);
// Inv#2:amt=Post(token[ast][this])-Pre(token[ast][this])
    }
  emit Deposit(vault, ast, safeAmt);
// Inv#3:safeAmt=Post(token[ast][msg.sender])-Pre(token[ast][msg.sender])
  }
}
\end{lstlisting}
\caption{The simplified deposit function.}
\label{fig:thorchain}
\end{figure}

\subsection{Challenges} \label{sec:motivation_challenges}

Smart Contract vulnerabilities associated with complex functionalities have led to numerous security incidents and occupy 81\% bountied vulnerabilities of Solidity-based
smart contracts from September 2021 to May 2023~\cite{wang2024smartinv}. These vulnerabilities usually require application-specific oracles for detection~\cite{zhang2023demystifying}. Fig.~\ref{fig:thorchain} illustrates a simplified function \texttt{\small deposit} in \textit{thorchain} contract~\footnote{0xc145990e84155416144c532e31f89b840ca8c2ce}. Normally, when a user invokes this function, the contract will transfer user's tokens to itself by executing \texttt{\small transferFrom} (line 6 or 10) and emits a deposit certificate, i.e., the event \texttt{\small Deposit} (line 13). In particular, the function \texttt{\small deposit} has two conditions (corresponding to two branches):

\begin{enumerate}
    \item \texttt{\small asset == RUNE}: the \texttt{\small asset} tokens are transferred to \texttt{\small vault} (line 6, \textit{Branch\#1}).
    \item \texttt{\small asset != RUNE}: the \texttt{\small asset} tokens are transferred to \texttt{\small \textcolor{blue}{this}} (line 10, \textit{Branch\#2}).
\end{enumerate}

However, since the contract lacks verification of whether tokens are actually deposited, the attacker could perform a ``fake deposit'' attack. Specifically, the attacker first deploys a fake token contract as \textit{fake\_token}, which does not transfer any tokens in function \texttt{\small transferFrom}. The attacker then calls the function \texttt{\small deposit} with the parameter \texttt{\small asset} = \textit{fake\_token}. As a result, the contract would normally execute \textit{Branch \#2} (line 8-10) and emit an event \texttt{\small Deposit} (line 13), and the attacker would obtain the deposit certificate without paying any tokens. 

To detect the above vulnerability, it is essential to understand the deposit logic and design specific oracles, that is, verifying ``the amount of tokens deposited is equal to the amount claimed by the user''. Unfortunately, these application-specific oracles always require a great deal of expert knowledge and a significant human effect in the design. Consequently, both the variety of vulnerabilities and the limitations of existing oracles pose significant challenges in protecting smart contracts from the harm of vulnerabilities.


\subsection{Our Insights} \label{sec:motivation_insight}

\noindent\textbf{Invariant and Vulnerability Detection.} We notice that smart contracts should strictly follow some invariants, which are regular rules for their functionalities~\cite{liu2022invcon,wang2024smartinv}. As shown in Fig.~\ref{fig:thorchain}, when users calls the function \texttt{\small deposit}, they should pay \texttt{\small safeAmt} amount of \texttt{\small ast} tokens, the function satisfies function invariant \textit{Inv\#3} ($token[ast][msg.sender]$ means the $ast$ token balance of $msg.sender$). Compared to function invariant, branch invariant is more granular and matches the functionality of the contract more accurately. For example, in Fig.~\ref{fig:thorchain}, \textit{Branch\#1} and \textit{Branch\#2} execute different functionalities (i.e., sending tokens to different addresses), so they satisfy different branch invariants, \textit{Inv\#1} and \textit{Inv\#2}, respectively. However, since the attacker does not pay any tokens to the contract when invoking the function \texttt{\small deposit}, the branch invariant \textit{Inv\#2} is violated. 

In fact, some existing work has used invariants as oracles, which verifies the satisfaction / violation of the invariant for vulnerability detection~\cite{duan2022towards, permenev2020verx}. However, their requirement of user-specific invariant limits their availability. 

\noindent\textbf{Invariant from Transactions.} Previous works~\cite{liu2022invcon, liu2022finding} prove that historical transactions contain abundant information about the functionalities of smart contracts. To obtain smart contract invariants, we have gained insight by observing the transactions in the case mentioned above. In benign transactions calling the function \texttt{\small deposit}, the parameter \texttt{\small amt} and the token balance always satisfy some properties (e.g., \textit{Inv\#3}), we could summarize these hiding properties as \textit{likely} invariants~\cite{ernst2007daikon}. In particular, since some functionalities are distinguished by different branches, the corresponding invariant is only satisfied in a certain branch (e.g., \textit{Inv\#1,\#2}). In these cases, the techniques that focus only on function or contract invariants (e.g., InvCon~\cite{liu2022invcon}) may fail to detect effective invariants.

To this end, our goal is to use dynamic invariant detection to extract fine-grained \textit{likely} invariants to approximate real invariants, which serve as application-specific oracles for vulnerability detection. 
	\section{SmartOracle} \label{sec:appraoch}

\begin{figure}[t]
    \centering
    \setlength{\abovecaptionskip}{0.05cm}
    \includegraphics[width=\linewidth]{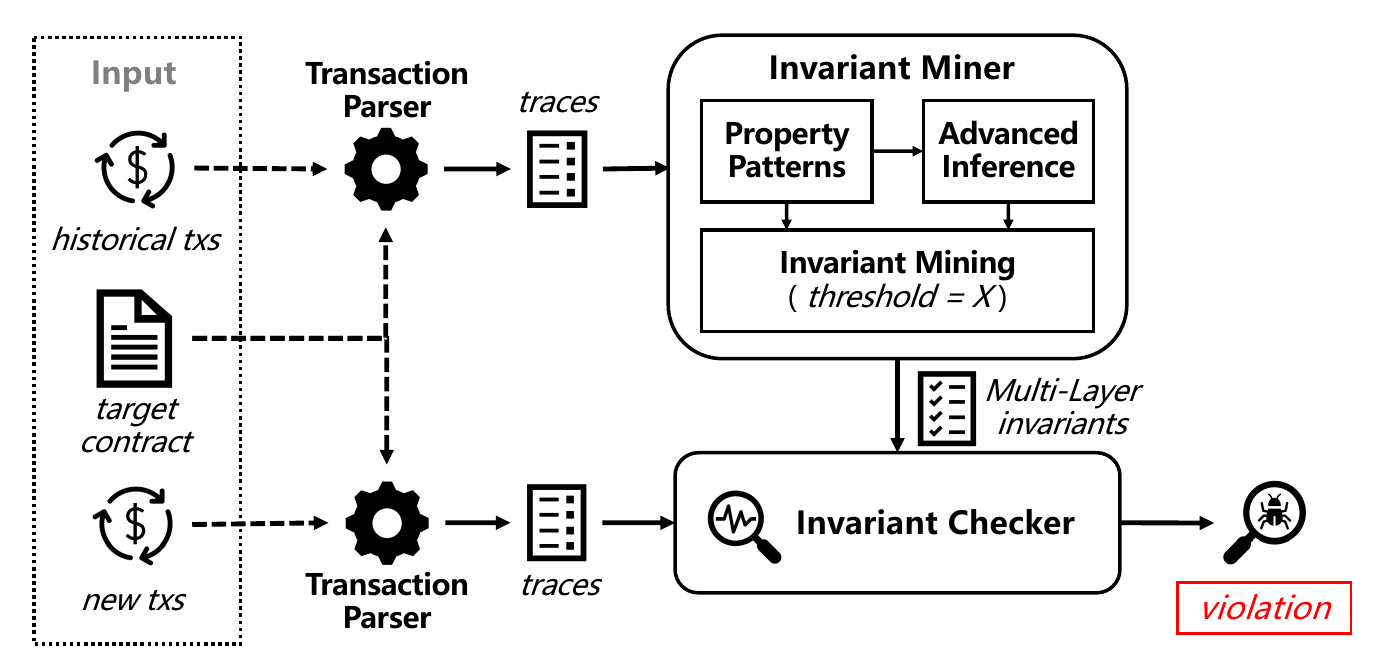}
    \caption{Overview of SmartOracle. SmartOracle mines invariants from \textit{historical txs} and detects the violations of invariants in \textit{new txs}.}
    \label{fig:overview}
\end{figure}

We propose SmartOracle, it automatically mines fine-grained \textit{likely} invariants for detecting smart contract vulnerabilities. As shown in Fig.~\ref{fig:overview}, SmartOracle consists of three components, i.e., \textit{Transaction Parser}, \textit{Invariant Miner} and \textit{Invariant Checker}. Given a target contract and its historical transactions (\textit{historical txs}), \textit{Transaction Parser} extracts the ABI and storage layout of the target contract, based on which it parses \textit{historical txs} and extracts their execution traces (\textit{traces}). To better fit the complicated contract functionalities, \textit{Invariant Miner} mine invariants of three layers, \textit{Contract}, \textit{Function}, and \textit{Branch}. For each layer, it employs pattern-based property detection and advanced property inference to construct properties, based on which it reserves the properties that satisfy $threshold$ of transactions as \textit{likely} invariants efficiently. Finally, given new transactions (\textit{new txs}), \textit{Invariant Checker} detects the transactions that violate the mined \textit{likely} invariants to detect vulnerabilities in the target contract.

\subsection{Transaction Parser}

In this stage, we aim to extract the execution traces of transactions. Since both \textit{historcial txs} and \textit{new txs} need to be parsed into execution traces for further analysis, we use \textit{transaction} uniformly to describe their extraction procedure for simplicity. As presented in Fig.~\ref{fig:tx_parser}, transaction parser extracts the ABI and the storage layout of the contract, based on which it parses the transactions and extracts their execution traces (\textit{trace}).

\subsubsection{Parsing Target Contract}
Given a smart contract, we utilize a static analysis framework called Slither~\cite{feist2019slither} to obtain its ABI and storage layout. The ABI records the name and type of the parameters in functions (events), which allows us to recover the value of parameters of functions (events) from their data in bytes. The storage layout records the name, type, location, and offset of storage slots for the state variable. It allows us to find state variables in storage slots and recover their value from bytes.

\subsubsection{Parsing Transactions}

Given a transaction of the target contract, we utilize our instrumented EVM to execute the raw transaction and record it as a call flow tree. Note that a call flow tree may contain multiple (internal) transactions derived from the external transaction. Therefore, we reserve the relevant transactions that call the target contracts and extract their execution traces as preparation for invariant detection.

\noindent\textbf{Instrumented EVM.}
In this part, instead of using the RPC \textit{debug\_traceTransaction}~\cite{debug_traceTransaction}, we instrument EVM to extract execution traces for higher efficiency~\cite{chen2019tokenscope}. Specifically, we insert our code into the function \textit{\small ApplyMessage()}~\cite{applymessage}, which is responsible for executing transactions and computing new states in EVM. In this way, a call stack is maintained to construct the tree structure of transactions and identify the transaction being executed. Furthermore, we insert recording codes into the handler of \texttt{CALL}, the way to initiate new transactions, to record the execution information of each transaction, such as emitted events and related account states.

\begin{figure}[t]
    \centering
    \setlength{\abovecaptionskip}{0.05cm}
    \includegraphics[width=\linewidth]{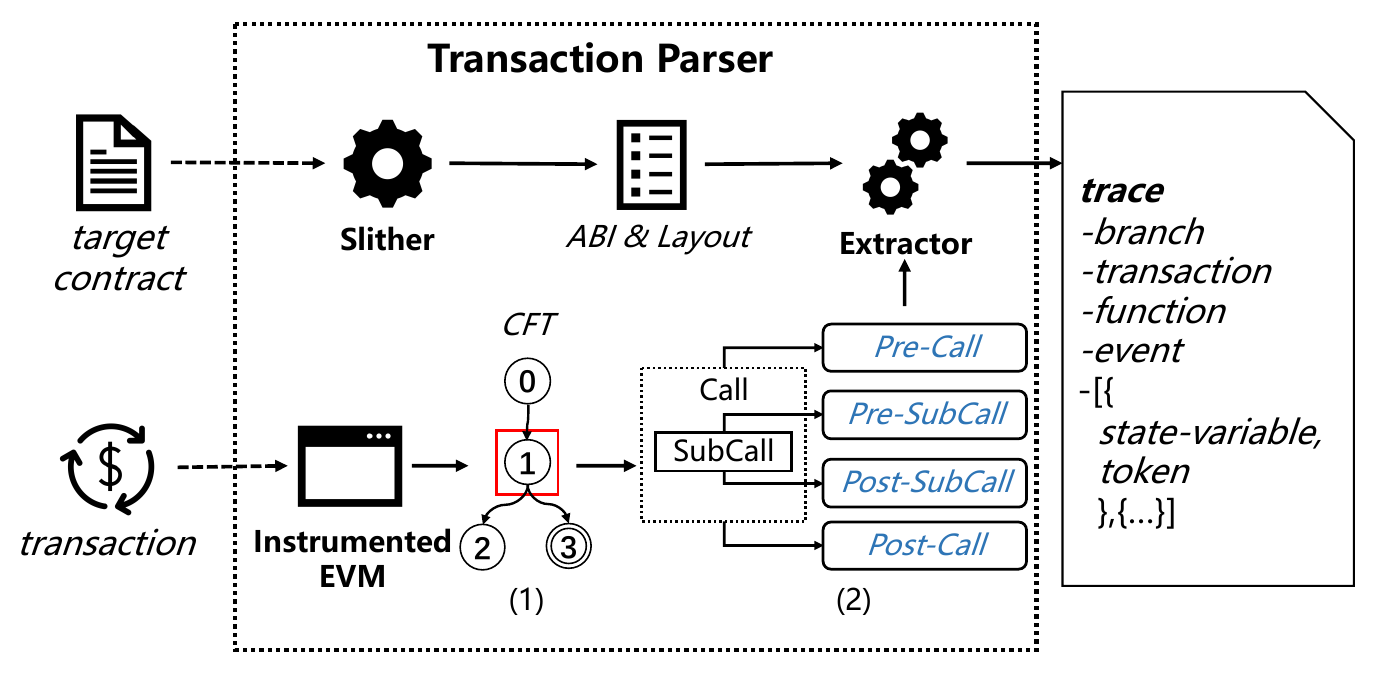}
    \caption{Workflow of Transaction Parser.}
    \label{fig:tx_parser}
\end{figure}

\noindent\textbf{Call Flow Tree (\textit{CFT}).} With the instrumented EVM, we execute the transaction to construct a tree structure that includes it and its derived internal transactions, as the \textit{CFT} shown in Fig.~\ref{fig:tx_parser}. Specifically, there are two types of nodes in \textit{CFT}, transaction nodes (e.g., \circnumber{1} in \textit{CFT}) and event nodes (e.g., \dcircnumber{3} in \textit{CFT}). The transaction node records the information of the call, including sender, receiver, the data recording the invoked function and its parameters, the amount of transferred ether, and related account states (i.e., the Ether balance and the storage slots being touched in the transaction). Furthermore, the transaction node also records the dynamic control flow of the call. The directed edge between the transaction nodes indicates that the parent transaction derives a new transaction. The event node includes the emitted address and parameters. The directed edge between the transaction node and the event node means that the transaction triggers the event. To filter out transactions relevant to the target contract, we traverse the \textit{CFT} and reserve the related subtrees in which the root transaction node calls the target contract. 

\subsubsection{Extracting Execution Trace}

In this part, we illustrate the extraction of execution traces based on the information gathered from contracts and transactions. For each subtree that calls the target contract, we first recover its execution branch from the dynamic control flow. Note that \texttt{JUMPI} is the only opcode that handles and changes the execution branches, so we record the destination of each \texttt{JUMPI} as \textit{branch}. Then we extract its trace and divide it into three categories, \textbf{Entry}, \textbf{Log} and \textbf{State}, which are shown in Table~\ref{tab:execution_trace}.

\noindent\textbf{(1) Entry.} Entry means the input of the contract call, which can be handled by the caller and represents the intention of the sender. It includes the information of the transaction and the invoked function.

(i) \textit{transaction} records the transaction information, such as the sender, receiver, block, and timestamp of the transaction.

(ii) \textit{function} records the function and its parameters. Based on the ABI of the target contract, we parse the bytecode of the transaction that records the signature and parameters of the invoked function to recover the value and types of parameters. 

\noindent\textbf{(2) Log.} Log records the logs (\textit{event}) defined by developers during contract execution. As a result, events include execution information (e.g., execution failed or succeeded) and indicate the behaviors of the contract. Similarly to \textit{function}, we use ABI to parse the bytecode of the event and recover the value and types of parameters. In particular, there may be multiple events within one transaction; we include all of them in the execution trace.

\begin{table}[t] 
    \centering
    \setlength{\abovecaptionskip}{0.05cm}
        \caption{Variables of execution trace.}
        \resizebox{\linewidth}{!}{
        \renewcommand{\arraystretch}{1.3}
        \begin{tabular}{l|l|l}
            \hline
            \textbf{Category} & \textbf{Variable} & \textbf{Description} \\
            \hline
            \multirow{2}{*}{\textbf{Entry}} & \textit{transaction} & sender, receiver, block, timestamp  \\
            & \textit{function} & the parameters of function \\
            \hline
            \textbf{Log} & \textit{event} & the information of emitted events \\
            \hline
            \multirow{2}{*}{\textbf{State}} & \textit{state-variable} & the value of state variables \\
            & \textit{token} & the value of token balances \\
            \hline
        \end{tabular}}
        \label{tab:execution_trace}
\end{table}

\noindent\textbf{(3) State.} State includes dynamic variables throughout the execution of the contract, representing the dynamic state of the contract. Since dynamic variables may change during the execution of the transaction, we record them in some essential \textit{program points}~\cite{ernst2007daikon}. On the one hand, the point that before and after the transaction execution is widely used in dynamic invariant analysis~\cite{liu2022invcon}. On the other hand, it is of great significance to record the dynamic variables at the points that the contract initiates internal transactions calling external contracts. This is because external contracts may alter the dynamic state and further exploit vulnerabilities in the target contract. For example, the attacker could exploit a \textit{Reentrancy} vulnerability during the contract calling external contracts~\cite{zheng2023turn}. Consequently, we record the dynamic variables of the contract as part of the execution trace at these record points: (i) before and after the transaction execution, as the \textit{Pre-Call} and \textit{Post-Call} in the (2) of Fig.~\ref{fig:tx_parser}; and (ii) before and after the contract initiating internal transaction calling external contracts, as the \textit{Pre-SubCall} and \textit{Post-SubCall} in the (2) of Fig.~\ref{fig:tx_parser}. At each record point, we record these dynamic variables: \textit{state-variable} and \textit{token}.

(i) \textit{state-variable} records the state variables of the contract. In particular, we only record the state variables that are read or written during the execution of the transaction to improve the efficiency of the analysis. 

Note that the smart contract stores its state variables in \textit{storage} (a mapping from \textit{bytes32} keys to storage slots that record \textit{bytes32} values), which does not indicate their corresponding state variables. In other words, given the \textit{storage} of the contract, we cannot directly recover their state variables. To this end, according to the storage layout of the contract, we first build the relationships between state variables and storage slots. We then find the corresponding storage slots of state variables and recover their names, types, and values. Specifically, state variables with \textit{Value Types}~\cite{solidity} (e.g., \textit{address}, \textit{uint256}) are arranged in the storage slots in order. Therefore, we can directly read the values of the state variables in their storage slots. For state variables with \textit{Reference Types}~\cite{solidity} (e.g., dynamic array), we can calculate the hash value of the slot to obtain the start slot of the array and recover the value of the state variable. For state variables with \textit{Mapping Types}~\cite{solidity}, they are stored in special storage slots (i.e., the hash value of the mapping key and the slot~\cite{ethereum}). In these cases, we first collect interesting values (e.g., the sender, the parameters of invoked function) as mapping keys. Then, we use them to calculate the hash values by \textit{keccak256} to locate the corresponding storage slots of state variables, and recover their value and types.

(ii) \textit{token} records the balances of Ether and ERC20 tokens, which represent the financial properties of the contract. For Ether, we directly read the balances of related addresses from their account states. For ERC20 tokens, we obtain the token balances of the related addresses by executing the \texttt{\small balanceOf} function in ERC20 token contracts. For convenience, we record \textit{token} variables as \textit{Mapping Types}, that is, \textit{token[tokenAddress][user]}, where \textit{tokenAddress} means the address of the token contract and \textit{user} means the address of the user. Similarly to \textit{state variable}, we reserve only token balances that are read or written during transaction execution to improve the efficiency of the analysis.

\subsection{Invariant Miner} \label{sec:inv_miner}

Based on the execution traces of transactions, we utilize an effective and efficient invariant detection to mine fine-grained smart contract invariants. 

\noindent\textbf{Multi-Layer Invariant.}
As discussed in Section~\ref{sec:motivation_insight}, a more fine-grained layer (i.e., branch layer) can accommodate more invariants of contract functionalities. However, due to the limited number of transactions for some branches (or functions), there may not be enough transactions to build confident invariants. In these cases, we use coarser invariants to fill the gap and ensure a comprehensive fit with contract functionalities. As in the example in Fig.~\ref{fig:thorchain}, if the transaction only invokes the \texttt{\small deposit} function to execute \textit{Branch\#1}, we cannot extract any invariants about \textit{Branch\#2} (e.g., \textit{Inv\#2}). Nonetheless, it is still possible to mine function invariants (e.g., \textit{Inv\#3}) as oracles.

Therefore, we divide execution traces into three layer groups: (1) Branch-layer group contains the traces of the same branch in a function; (2) Function-layer group contains the traces of the same function; (3) Contract-layer group contains all traces of the contract. 

Consequently, we perform an invariant mining on each group to obtain its corresponding invariants. Specifically, we perform pattern-based detection and advanced inference to construct properties, and reserve the properties that satisfy across transactions as \textit{likely} invariants. Furthermore, we propose a heuristic threshold-based mining strategy to increase the usefulness of invariant mining.

\subsubsection{Property Construction}


For each execution trace, we first use a pattern-based property detection to obtain \textit{basic properties}. Then, we perform advanced property inference to derive more \textit{advanced properties}. Both serve as candidate invariants for the final \textit{likely} invariants.

\noindent\textbf{Pattern-based Detection.}
We design property patterns of variables based on the specific characteristics of smart contracts, then adapt them to the trace variables and obtain the candidate \textit{basic invariants}. Next, we illustrate the main patterns as follows ($x$, $y$ and $z$ denote variables; $a$ and $b$ refer to calculated constants).

(i) \textit{Comparison} pattern refers to common properties among variables, such as $x = y$, $x = -y$, and $x == a$. These patterns are prevalent in smart contracts. 

(ii) \textit{Membership} pattern means that a variable is a member of another variable. Considering $y$ is an \textit{Array} variable that can include elements in \textit{Value Types} (e.g., uint256) or \textit{Struct}, we denote $x \in y$ if one of the following conditions is satisfied: (i) if the member of $y$ is \textit{Value Types}, $x$ is equal to a member of $y$; (ii) if the member of $y$ is \textit{Struct}, $x$ is equal to one of the components of a member of $y$ (e.g., $y[0].address = x$). This pattern also occurs frequently in the smart contracts with access control checks. For example, some functions implement access control through a whitelist, so they hold the property like \texttt{\small msg.sender $\in$ whitelist}.

(iii) \textit{Arithmetic} pattern describes the arithmetic relations between variables, including linear relations (e.g., $z = ax +by + c$) and quadratic relations (e.g., $xy = a$). In particular, DeFi apps always use these arithmetic relations to carry out their financial models. For example, Uniswap V2~\cite{uniswapv2} performs its swap business based on the relation $xy = a$. Thus, using these arithmetic patterns can effectively model the relation of variables in financial models.

\noindent\textbf{Advanced Inference.}
Based on the \textit{basic properties}, we further derive them to obtain \textit{advanced properties}, which are more abstract and fit more complicated functionalities. Precisely, we infer new properties based on properties containing \textit{Mapping Types} variables, which are essential for implementing contract functionalities. We use the example in Fig.~\ref{fig:thorchain} to explain the inference procedure. Assume that \textit{Comparison} pattern constructs a property within one execution trace:
\begin{align}
    \begin{split} 
        saftAmt &= Post(token[addr\_A][addr\_B]) \\
        & -Pre(token[addr\_A][addr\_B]) \\
    \end{split} 
\end{align}
where $addr\_A$ and $addr\_B$ are the values of \textcolor{teal}{address} type. And, the variables in the execution trace satisfy:
\begin{align}
    \begin{split} 
        ast = addr\_A; \quad msg.sender = addr\_B
    \end{split} 
\end{align}
In this case, the \textit{Comparsion} property (1) only represents the relation between $safeAmt$ and a normal token balance. Given the variables in (2), we could infer that the $addr\_A$ and $addr\_B$ in property (1) are decided by the parameter \textit{ast} and the sender of transaction \textit{msg.sender}, and derive a new property as follows:
\begin{align}
    \begin{split} 
    saftAmt &= Post(token[ast][msg.sender]) \\
        & -Pre(token[ast][msg.sender]) \\
    \end{split} 
\end{align}
In this way, we replace a concrete value with a symbolic value to construct a new abstract property that may be suitable for more complicated contract functionalities. Furthermore, considering that some properties can be inferred many times, we iteratively perform this advanced inference on properties until no new properties are constructed.

\subsubsection{Mining Strategy}

Based on property construction, we heuristically propose a threshold mechanism to improve the robustness of invariant mining. 

\noindent\textbf{Threshold.} Normally, we should only reserve properties that satisfy all historical transactions as \textit{likely} invariants~\cite{liu2022invcon}. However, we cannot ensure that all historical transactions are normal; that is, the transaction is benign and executed normally without any exception. In contrast, abnormal cases hiding in historical transactions may interfere with mining invariants. For example, some contracts had been attacked due to the abnormal state maintained by the manager but are still active after being recovered, or only emit event \textit{Failure} but do not revert when an exception occurs. As a result, we may fail to recognize invariants to distinguish normal transactions from abnormal transactions (e.g., attacks) that trigger vulnerabilities. 

To this end, we heuristically use a hyperparameter $threshold \in (0,1]$ to increase the robustness of the mining strategy. Specifically, we decrease $threshold$ and consider properties that satisfy $threshold$ percentage of historical transactions as invariants, which can reduce the interference of hidden abnormal cases in historical transactions.

However, since decreasing $threshold$ means loss of confidence, SmartOracle may mine biased or false invariants and report false positives in detection results. Therefore, we adjust $threshold$ to suit different practical requirements. For example, we increase $threshold$ when we need more precise invariants as oracles.

\RestyleAlgo{ruled}
\IncMargin{0.1em}
\SetKwComment{Comment}{/* }{ */}
\begin{algorithm}[t]
\caption{Workflow of Invariant Mining.}\label{alg:inv_mining}
\SetKwInOut{Input}{input}\SetKwInOut{Output}{output}
\Input{$traces$, $threshold$}
\Output{$invariants$}
\BlankLine
\Comment{$Property \quad Searching$}
$candiInvs := mapping()$\;
\For {$tr$ in $traces$}{
    $basicProperties$ := $patternDetection(tr)$\;
    \For{$basicProperty$ in $basicProperties$}{
        $candiInvs[basicProperty] += 1$\;
        $advancedProperties$ := []\;
        $oldPs$ := $[basicProperty]$\;
        \For{$len(oldPs) > 0$}{
            $newPs$ := $inference(oldPs, tr)$\;
            $oldPs$ = $newPs$\;
            $advancedProperties.extend(newPs)$\;
        }
        \For{$property$ in $advancedProperties$}{
            $candiInvs[property] += 1$\;
        }
    }
}
\Comment{$Invariant \quad Filtering$}
\For {$inv$ in $candiInvs$}{
    \If{$candiInvs[inv] < threshold * len(traces)$}{
        $candiInvs.pop(inv)$\;
    }
}
$invariants := remoevRedundant(candiInvs)$\;
\end{algorithm}

\noindent\textbf{Algorithm.} We illustrate the invariant mining workflow as described in Algorithm~\ref{alg:inv_mining}. The key procedures are as follows: 


\begin{enumerate}
    \item \textit{Property Searching}. For each trace in $traces$, we use pattern-based property detection to construct \textit{basicProperties}. For each property in \textit{basicProperties}, we perform an iterative advanced inference to obtain \textit{advancedProperties}. Both serve as candidate invariants (\textit{candiInvs}).
    \item \textit{Property Filtering}. For each candidate invariant, we count the number of traces that satisfy the invariant and discard the candidate invariants that cannot be satisfied by $threshold$ of traces. Finally, we remove redundant candidate invariants and set the remaining cases as \textit{likely} invariants. 
\end{enumerate}

In particular, since we solely search for properties for each transaction and reserve those that satisfy $threshold$ percentage of all historical transactions, this algorithm is deterministic without randomness.

\subsection{Invariant Checker}

Based on the mined \textit{likely} invariants (\textit{invariants}), SmartOracle detects smart contract vulnerabilities by hunting the violations of invariants within transactions. Given each transaction in \textit{new txs}, SmartOracle first parses the transactions to obtain its execution trace. Then SmartOracle finds the corresponding most fine-grained set of invariants ($checked\_invs$), the detailed procedure is as follows: (1) if there is the corresponding branch-layer invariant set, using it as $checked\_invs$; otherwise (2) if there is the corresponding function-layer invsariant set, using it as $checked\_invs$; otherwise (3) using the contract-layer invariant set as $checked\_invs$. Finally, SmartOracle checks the satisfaction of $checked\_invs$ for the trace and reports the violated invariants of the transaction.

        \section{Experiments} \label{sec:exps}

In this section, we conduct experiments to evaluate the effectiveness of SmartOracle and aim to answer the following research questions:

\begin{itemize}

\item RQ1: How effective is SmartOracle in mining smart contract invariants?
\item RQ2: How effective is SmartOracle in detecting smart contract vulnerabilities?
\item RQ3: What types of invariants does SmartOracle extract from smart contracts?
\item RQ4: How efficient is SmartOracle?

\end{itemize}

\subsection{RQ1: Effectiveness of Invariant Detection} \label{sec:effectiveness_mining}

To evaluate the effectiveness of invariant detection, we utilize a benchmark from InvCon~\cite{liu2022invcon}, which contains 246 ERC20 contracts and their invariants. Note that ERC20 contracts satisfy the ERC20 standard~\cite{erc20} and largely follow the same set of common invariants. Table~\ref{tab:erc20_invs} demonstrates the common ERC20 invariants~\cite{liu2022invcon}, representing the basic functionalities of the contract and its three functions (i.e., \textit{transfer}, \textit{approve}, and \textit{transferFrom}). For example, \textit{inv\#2} describes the relation between the function parameter \texttt{\small amt} and the state variable \texttt{\small balances} during token transfer. To this end, we apply SmartOracle to this benchmark and use these seven common invariants to evaluate the effectiveness of invariant detection. Furthermore, we run SmartOracle with the same input as InvCon to ensure a fair comparison. Table~\ref{tab:effectiveness_mining} shows the statistical results of the experiments, \#N means the number of detected \textit{likely} invariants, and \#TP means the number of invariants detected correctly. To ensure the correctness of the results, we perform manual verification of the detected invariants based on the labels in the benchmark.

\subsubsection{Ability of Invariant Detection}
Since InvCon only extracts invariants that are satisfied in all transactions, we also run SmartOracle with the corresponding hyperparameter $threshold=1$. Overall, SmartOracle detects at least one ERC20 invariant in 167 unique contracts, and a total of 518 \textit{likeyly} invariants (50\% ($\frac{518-256}{518}$) more than those detected by InvCon). For function \textit{transfer}, the \textit{likely} invariants detected by SmartOracle are much more than those detected by InvCon. This is mainly because InvCon only focuses on function invariants, while SmartOracle can handle more fine-grained invariants (i.e., branch invariants). As shown in Fig.~\ref{fig:erc20}, the ERC20 \textit{Inv\#3} is only satisfied for the branch with a specific condition (line 7) but not for the whole function so that this invariant can be detected by SmartOracle but not by InvCon. In addition, SmartOracle only detects a few \textit{Inv\#3,\#4,\#5} invariants due to the lack of transactions calling function \textit{transferFrom}. However, since many ERC20 contracts involve only very few historical transactions, SmartOracle cannot detect their ERC20 invariants.

\begin{table}[t] 
    \centering
    \setlength{\abovecaptionskip}{0.05cm}
        \caption{Common ERC20 invariants.}
        \resizebox{\linewidth}{!}{
        \renewcommand{\arraystretch}{1.3}
        \begin{tabular}{l|l|l}
            \hline
            & ID & Invariant \\
            \hline
            \textit{Contract} & \textit{Inv\#1} & $SUM(balances) = totalSupply$ \\
            \hline
            \multirow{3}{*}{\textit{transfer}} & \multirow{2}{*}{\textit{Inv\#2}} & $amt = Pre(balances[msg.sender])$ \\
            & &$ \quad -Post(balances[msg.sender])$ \\
            & \textit{Inv\#3} & $amt = Post(balances[to]) - Pre(balances[to])$ \\
            \hline
            \textit{approve} & \textit{Inv\#4} & $amt = Post(allowance[msg.sender][spender])$ \\
            \hline
            \multirow{4}{*}{\textit{transferFrom}} &\multirow{2}{*}{\textit{Inv\#5}} & $amt = Post(allowance[from][msg.sender])$ \\
            & & $ \quad - Pre(allowance[from][msg.sender])$ \\
            & \textit{Inv\#6} & $amt = Pre(balances[from]) - Post(balances[from])$ \\
            & \textit{Inv\#7} & $amt = Post(balances[to]) - Pre(balances[to])$ \\
            \hline
        \end{tabular}}
        \label{tab:erc20_invs}
\end{table}

\begin{table}[t] 
    \centering
    \setlength{\abovecaptionskip}{0.05cm}
        \caption{Statistics about the detected ERC20 invariants.}
        \resizebox{\linewidth}{!}{
        \renewcommand{\arraystretch}{1.3}
        \begin{tabular}{l|rr|rr|rr|rr}
            \hline
            \multirow{3}{*}{\textbf{Inv} (num.)} & \multicolumn{2}{c|}{\multirow{2}{*}{\textbf{InvCon}}} & \multicolumn{6}{c}{\textbf{SmartOracle}} \\
            & & & \multicolumn{2}{c|}{\textit{threshold=1.0}} & \multicolumn{2}{c|}{\textit{threshold=0.9}} & \multicolumn{2}{c}{\textit{threshold=0.8}} \\
            & \#TP & \#N & \#TP & \#N & \#TP & \#N & \#TP & \#N \\
            \hline
\textit{Inv\#1} (233) & \textbf{121} & 123 & \textbf{126} & 126 & 126 & 126 & \textbf{126} & 126 \\
\textit{Inv\#2} (225) & \textbf{39} & 45 & \textbf{160} & 169 & 169 & 181 & \textbf{170} & 183 \\
\textit{Inv\#3} (225) & \textbf{20} & 22 & \textbf{125} & 134 & 132 & 143 & \textbf{133} & 145 \\
\textit{Inv\#4} (244) & \textbf{49} & 50 & \textbf{58} & 58 & 62 & 62 & \textbf{63} & 63 \\
\textit{Inv\#5} (228) & \textbf{2} & 2 & \textbf{9} & 9 & 9 & 9 & \textbf{9} & 9 \\
\textit{Inv\#6} (219) & \textbf{8} & 9 & \textbf{12} & 13 & 12 & 13 & \textbf{12} & 13 \\
\textit{Inv\#7} (219) & \textbf{4} & 5 & \textbf{9} & 9 & 9 & 9 & \textbf{9} & 9 \\
\hline
\textit{All} & \textbf{243} & 256 & \textbf{499} & 518 & 519 & 543 & \textbf{522} & 548 \\
\hline
        \end{tabular}}
        \label{tab:effectiveness_mining}
\end{table}

\subsubsection{Impact of Threshold}
To evaluate the impact of \textit{threshold} in mining \textit{likely} invariants, we run SmartOracle with hyperparameter $(threshold = 1.0, 0.9, 0.8)$ and show the results in Table~\ref{tab:effectiveness_mining}. As the \textit{threshold} decreases from $1.0$ to $0.8$, the confidence of \textit{likely} invariants decreases, and the precision rate of detected \textit{likely} invariants slightly decreases from 96\% to 95\%. However, the total number of true detected invariants increases by approximately 5\% ($\frac{522-499}{499}$). Therefore, slightly decreasing the \textit{threshold} can effectively detect more invariants with an acceptable loss of precision. These experimental results demonstrate the usefulness of our design threshold mechanism in practice applications.

\subsubsection{Impact of Transactions}

\begin{figure}[t]
    \centering
    \setlength{\abovecaptionskip}{0.05cm}
    \includegraphics[width=\linewidth]{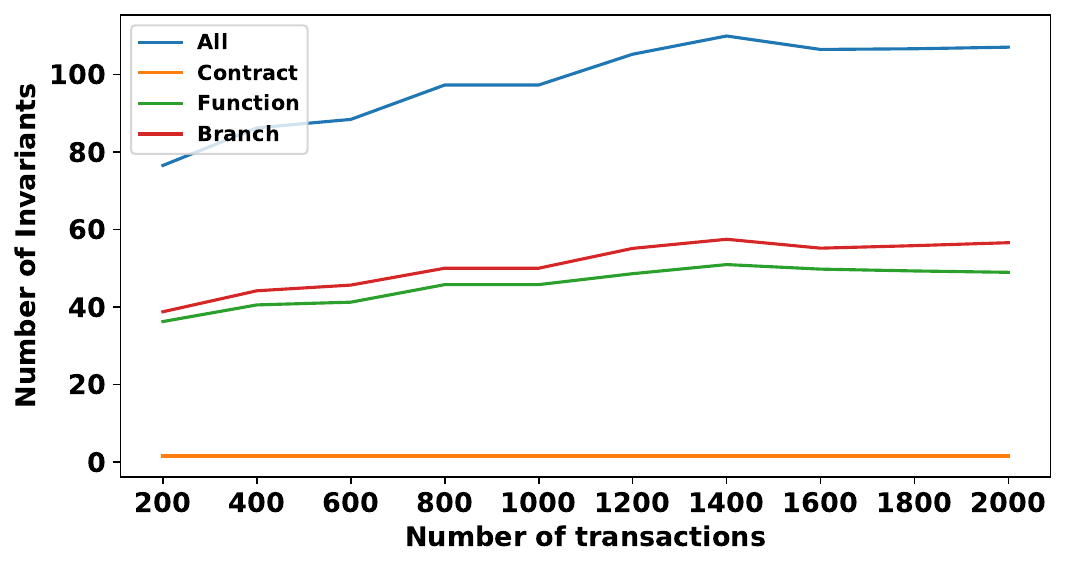}
    \caption{Average number of detected invariants along transactions.}
    \label{fig:tx_num_impact}
\end{figure}

As SmartOracle mines \textit{likely} invariants from transactions, the number of transactions may affect its effectiveness. To investigate the impact, we select 17 ERC20 contracts that involve more than 2,000 transactions, ensuring sufficient data for further evaluation. With \textit{threshold=1}, we use SmartOracle to mine invariants from the earliest $(N=200, ... , 2000)$ transactions of contracts, the experimental results are shown in Fig.~\ref{fig:tx_num_impact}.

The total number of invariants and the number of invariants in each layer show the same growth trend. Initially, these few transactions (e.g., 200) cover only a few functionalities. As the amount of transactions increases, the mined number of mined invariants increases rapidly and eventually levels off. After reaching 1400, the number of invariants decreases slightly as more transactions enable SmartOracle to remove biased or false invariants. In particular, the number of branch invariants is 7-13\% more than that of function invariants, indicating that branch invariants can accommodate more fine-grained contract properties.

\subsection{RQ2: Effectiveness of Vulnerability Detection} \label{sec:effectiveness}

To detect smart contract vulnerabilities with invariants, we need to mine the essential invariants. These invariants should be satisfied with respect to benign behavior, but their violation implies malicious behavior (e.g., triggering vulnerabilities). First, we build a benchmark that contains several real-world vulnerabilities. Then, we apply SmartOracle to this benchmark and evaluate its effectiveness in mining essential invariants for vulnerability detection.

\subsubsection{Dataset} We reference and expand on an existing dataset~\cite{su2023defiwarder} by investigating real-world security incidents on Ethereum. Specifically, we collect the security incidents between January 2020 and June 2022 in two well-known online libraries, namely Slowmist~\cite{slowmisthacked} and Rekt~\cite{rekt}. From these sources, we exclude incidents related to social engineering (e.g., rug pull and private-key leak) and focus solely on smart contracts that were attacked due to vulnerabilities.

Our dataset contains 65 vulnerable smart contracts from real-world DApps, with an average of 1,973 lines of code (LOC) and associated with over 1.6 million related transactions. According to previous work~\cite{wang2024smartinv}, we manually classify these vulnerabilities, including 7 \textit{Reentrancy} (RE), 11 \textit{Privilege Escalation} (PE), 17 \textit{Price Manipulation} (PM) and 30 \textit{Business Logic Flaw} (BLF).


\subsubsection{Experimental Setup}
We divide the related transactions of the contract into two parts: (i) transactions that occurred before the attack (i.e., \textit{historical txs}) and (ii) transactions of attacks and the subsequent 1,000 transactions (i.e., \textit{new txs}). Given \textit{historical txs}, we use SmartOracle to mine \textit{likely} invariants with hyperparameter $(threshold = 0.90, 0.92, ..., 1.00)$. Then, SmartOracle utilizes mined invariants to reveal the transactions violating invariants in \textit{new txs} (a total of 14,973 transactions). 

\subsubsection{Evaluation of SmartOracle}
The experimental results are shown in Table~\ref{tab:effectiveness}. For transactions that violate mined invariants (namely \textit{violated transactions}), we mark those indeed caused by vulnerabilities as True Positive (\textit{TP}); otherwise, mark them as False Positive (\textit{FP}). In terms of smart contracts, if a contract contains \textit{TP} violated transactions, we label the contract as \textit{TP}, and vice versa. Particularly, if SmartOracle detects \textit{TP} and \textit{FP} violated transactions in the contract at the same time, we mark it as \textit{TP\&FP}.

\begin{table}[t] 
    \centering
    \setlength{\abovecaptionskip}{0.05cm}
        \caption{The results of SmartOracle with different $threshold$.}
        \resizebox{0.85\linewidth}{!}{
        \renewcommand{\arraystretch}{1.3}
        \begin{tabular}{c|ccc|rr}
            \hline
            \multirow{2}{*}{\textit{threshold}} & \multicolumn{3}{c|}{\textbf{Contract}} & \multicolumn{2}{c}{\textbf{Transaction}} \\
            & \textbf{\#TPs} & \textbf{\#TP\&FPs} & \textbf{\#FPs} & \textbf{\#TPs} & \textbf{\#FPs} \\
            \hline
0.90 & 36 & 10 & 13 & 575 & 326 \\
0.92 & 36 & 8 & 11 & 575 & 272 \\
0.94 & 34 & 6 & 10 & 555 & 62 \\
0.96 & 32 & 4 & 7 & 515 & 24 \\
\textbf{0.98} & \textbf{31} & \textbf{3} & \textbf{5} & \textbf{466} & \textbf{17} \\
1.00 & 29 & 3 & 5 & 312 & 15 \\
            \hline
        \end{tabular}}
        \label{tab:effectiveness}
\end{table}

\noindent\textbf{Impact of Threshold.} When $threshold = 0.90$, SmartOracle reports many \textit{FP} violated transactions since SmartOracle mines biased invariants that lead to false positives. As the hyperparameter $threshold$ increases from $0.90$ to $0.98$, the confidence of invariants increases, and the number of false positives decreases. In particular, increasing $threshold$ also results in a decrease in true positives, since abnormal cases in \textit{historical txs} interfere with the mining invariants of normal transactions. Fortunately, compared to false positives (62\% = $\frac{13-5}{13}$), the reduction of true positives (14\% = $\frac{36-31}{36}$) is worth and acceptable.

\noindent\textbf{True Positive.} In the context of $threshold = 0.98$, SmartOracle totally detects 483 violated transactions, 96\% ($\frac{466}{483}$) of which are true positives (i.e., caused by smart contract vulnerabilities). Regarding contracts, SmartOracle successfully reveals true vulnerabilities in 31 vulnerable contracts, 28 of which contain only \textit{TP} violated transactions. Moreover, as shown in Table~\ref{tab:other_tools_effectiveness}, most detected vulnerabilities are associated with contract functionalities, such as 6 PM, 6 PE, and 14 BLF vulnerabilities. These results prove the feasibility of mining \textit{likely} invariants from historical transactions as application-specific oracles for vulnerability detection. SmartOracle can effectively detect smart contract vulnerabilities with an acceptable precision rate.

\noindent\textbf{False Positive.} SmartOracle also reports 5 \textit{FP} vulnerable contracts. However, these vulnerable contracts only involve 17 \textit{FP} violated transactions. After manual analysis, we find that these violated transactions are mainly caused by atypical scenarios within the smart contracts. These scenarios are benign but have not occurred in \textit{historical txs}, which makes SmartOracle mine biased invariants and report false positives.
Fortunately, these atypical cases are easily distinguished from \textit{TP} violated transactions with very little manual effect.

\noindent\textbf{False Negative.} However, there are still some vulnerable contracts that SmartOracle cannot detect. We summarize the main reasons as follows: (1) before being attacked, the vulnerable function had been invoked with too few transactions (even no transaction) that cover limited functionalities, so SmartOracle could not mine effective invariants for detection; (2) the smart contract vulnerabilities violate the invariants that involve the variables across multiple contracts. For example, exploiting some \textit{Price Manipulation} requires manipulating the state variables in other contracts (e.g., the variables in price oracle contract~\cite{kong2023defitainter}). In these cases, SmartOracle cannot handle the variables in other contracts and mine the corresponding invariant for vulnerability detection. Since handling multiple contracts would involve significant computational overhead on SmartOracle, we do not deal with this kind of case in this work.

\begin{table}[t] 
    \centering
    \setlength{\abovecaptionskip}{0.05cm}
        \caption{The experimental results of evaluating SmartOracle and other tools on our benchmark.}
        \resizebox{\linewidth}{!}{
        \renewcommand{\arraystretch}{1.3}
        \begin{tabular}{l|cc|cc|cc|c}
            \hline
            \multirow{2}{*}{\textbf{Tool}} & \multicolumn{2}{c|}{\textbf{PE (11)}} & \multicolumn{2}{c|}{\textbf{PM (17)}} & \multicolumn{2}{c|}{\textbf{RE (7)}} & \textbf{BLF (30) } \\
            & \textbf{\#TPs} & \textbf{\#FPs} & \textbf{\#TPs} & \textbf{\#FPs} & \textbf{\#TPs} & \textbf{\#FPs} & \textbf{\#TPs}\\
            \hline
            SmartOracle & 6 & / & 6 & / & 5 & / & 14 \\
            \hline
            AChecker & 0 & 0 & / & / &  / & / & / \\
            \hline
            DeFiTainter & / & / & 6 & 0 & / & / & / \\
            \hline
            Slither & / & / & / & / & 4 & 39 & / \\
            Sailfish & / & / & / & / & 1 & 8 & / \\
            \hline
        \end{tabular}}
        \label{tab:other_tools_effectiveness}
\end{table}

\subsubsection{Comparing SmartOracle with existing tools}
We also adapt existing tools to our collected dataset for comparison. Based on the vulnerability types in the vulnerable contracts, we select Slither~\cite{feist2019slither}, Sailfish~\cite{bose2022sailfish} for \textit{Reentrancy},  AChecker~\cite{ghaleb2023achecker} for \textit{Privilege Escalation}, and DeFiTainter~\cite{kong2023defitainter} for \textit{Price Manipulation}. For 65 vulnerable contracts, we reference previous work~\cite{su2023defiwarder} and set the analysis timeout of tools to 30 minutes for each smart contract. SmartOracle performs an analysis with $threshold=0.98$. Table~\ref{tab:other_tools_effectiveness} shows the detection results; True positive (TP) means that the tool detects the true vulnerability, while False positive (FP) means that the tool reports the vulnerability that the contract does not contain.

For \textit{Reentrancy}, Slither reveals 4 true positives but reports 39 false positives, while Sailfish detects 1 true positive alongside 8 false positives. The high false-positive rates question their practicality in real-world applications. For \textit{Privilege Escalation}, AChecker fails to reveal any vulnerabilities. It suffers from out-of-memory errors in many contracts, indicating that AChecker is impractical in real-world complex DApp contracts (e.g., with thousands of lines of code). For \textit{Price Manipulation}, SmartOracle is close to DeFiTainter in effectiveness. 
In addition to the above three vulnerabilities, SmartOracle detects other 14 \textit{Business Logic Flaw} vulnerabilities, covering a broader scope of vulnerabilities in real-world security incidents. Overall, existing tools have limitations in detecting smart contract vulnerabilities in real-world DApps, while SmartOracle can provide more effective protection by run-time verification.

\subsection{RQ3: Invariants of Smart Contracts}

In this experiment, we study the distributions of the smart contract invariants extracted by SmartOracle. We count the proportion of variables (construction ways) in invariants mined from \textit{historical txs} (namely \textit{mined invariants}) and invariants violated in \textit{new txs} (namely \textit{violated invariants}), respectively.

\begin{figure}[t]
    \centering
    \setlength{\abovecaptionskip}{0.05cm}
    \includegraphics[width=\linewidth]{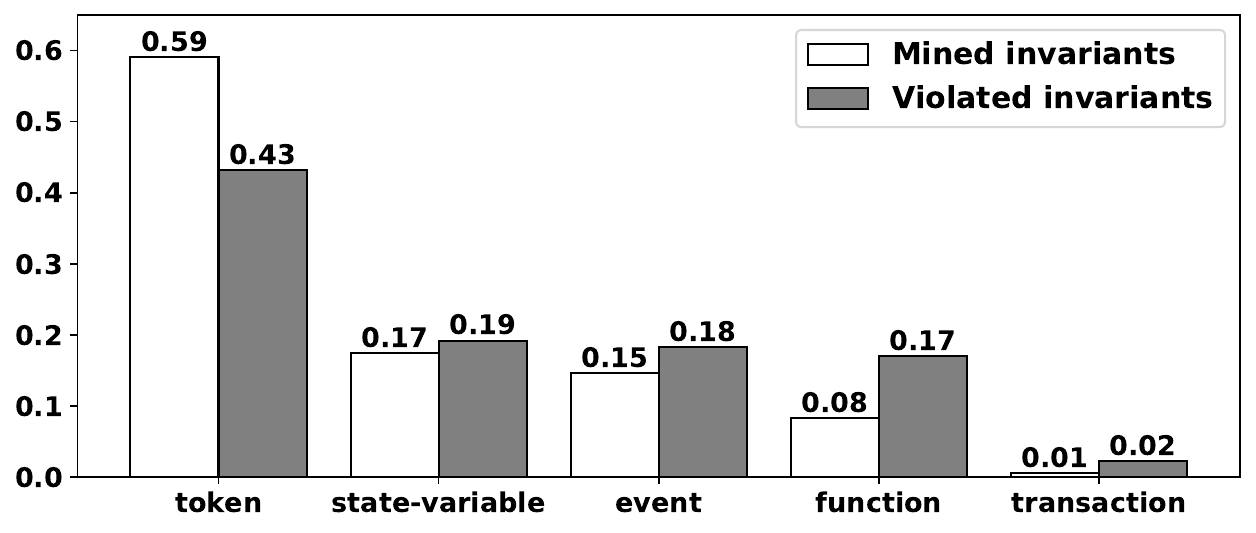}
    \caption{Proportions of variables in the invariants mined from historical transactions (mined invariants) and the invariants violated in new transactions (violated invariants).}
    \label{fig:inv_variable}
\end{figure}

\begin{figure}[t]
    \centering
    \setlength{\abovecaptionskip}{0.05cm}
    \includegraphics[width=\linewidth]{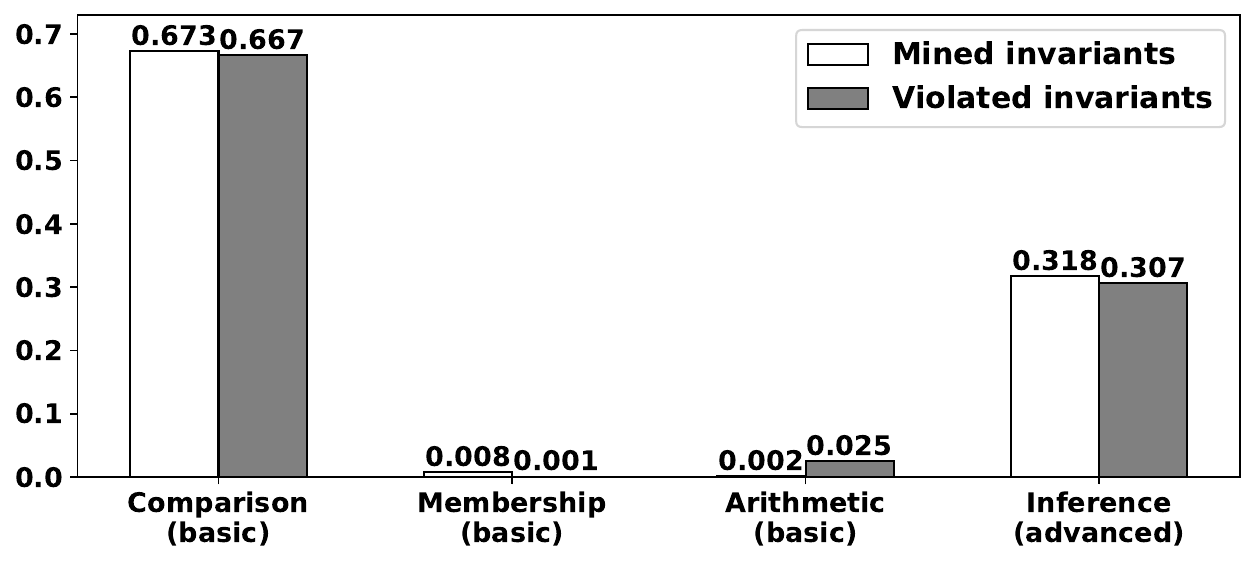}
    \caption{Proportions of construction in the invariants mined from historical transactions (mined invariants) and the invariants violated in new transactions (violated invariants).}
    \label{fig:inv_pattern}
\end{figure}

\noindent\textbf{Variables in invariants.} Fig.~\ref{fig:inv_variable} shows the proportion of each variable (listed in Table~\ref{tab:execution_trace}) in invariants extracted by SmartOracle. Notably, invariants involving \textit{token} variables constitute the highest proportion in both mined invariants (59\%) and violated invariants (43\%). This is because \textit{token} variable is the essential element of many business operations on Ethereum, especially of common financial servers.
About 19\% of the mined and violated invariants involve \textit{state-variable}, which are the basis of smart contracts to record states and participate in most functionalities of smart contracts. In addition, \textit{function} and \textit{event} variables occupy similar proportions in the violated invariants.


\noindent\textbf{Construction of invariants.} Fig.~\ref{fig:inv_pattern} shows the proportion of invariant construction extracted by SmartOracle. We find that each construction has a similar proportion between the mined invariants and the violated invariants. 
For the basic invariants constructed by pattern-based detection, the invariants involving \textit{Comparison} occupy the most significant proportion since \textit{Comparison} is the most common property in smart contracts. However, since there are very few \textit{Array} variables in smart contracts, \textit{Membership} only constructs very few invariants. In addition, \textit{Arithmetic} pattern contains the minimal part of the invariants, but the invariants still successfully reveal some smart contract vulnerabilities. \textit{Inference} means the invariants constructed by advanced inference; it occupies approximately 30\% of all invariants, indicating the great effectiveness of our designed advanced inference in constructing invariants.

\subsection{RQ4: Efficiency of SmartOracle}

\begin{table}[t] 
    \centering
    \setlength{\abovecaptionskip}{0.05cm}
        \caption{The average time consumption of analyzing per transaction.\textit{ Total Checking Time} is the sum of time consumption of \textit{Transaction Parser} and \textit{Invariant Checker}.}
        \resizebox{0.8\linewidth}{!}{
        \renewcommand{\arraystretch}{1.3}
        \begin{tabular}{l|l}
            \hline
            Action & Average Time ($ \times 10^{-3}$ second) \\
            \hline
            \textit{Transaction Parser} & 44.66 \\
            \textit{Invariant Miner} & 37.96 \\
            \textit{Invariant Checker} & 7.02 \\
            \hline 
            \textit{Total Checking Time} & 51.68 (44.66 + 7.02) \\
            \hline
        \end{tabular}}
        \label{tab:time_consumption}
\end{table}

In this RQ, we evaluate the efficiency of SmartOracle in real-time detection. Specifically, we perform SmartOracle on related transactions of smart contracts, involving 31,184 transactions for mining and 14,973 transactions for checking. We record the average execution time per transaction of each procedure, including \textit{Transaction Parser}, \textit{Invariant Miner}, and \textit{Invariant Checker}.

The experimental results are presented in Table~\ref{tab:time_consumption}. We only carry out \textit{Invariant Miner} once for each smart contract, then use the mined invariants for bug detection. Therefore, in the checking scenario, the \textit{Total Checking Time} is the sum of the time consumption of \textit{Transaction Parser} and \textit{Invariant Checker}. SmartOracle checks each transaction with $51.68 \times 10^{-3}$ seconds in average. In particular, SmartOracle's TPS (19.3 = $\frac{1000}{51.68}$) is higher than that of Ethereum (approximately 12.0)~\cite{etherscan}, which means that SmartOracle has the ability to perform real-time audits and promptly reveal violated transactions as they occur. Furthermore, for bugs that require multiple transactions to be exploited, SmartOracle can immediately hunt the violated transactions before the whole exploit is finished. For example, the entire exploit of \textit{bZx} DApp~\cite{bzxhacked} involves multiple transactions, which span nearly 8 minutes from the first violated transaction detected by SmartOracle to the final profitable transaction. Therefore, if we could reveal the first violated transaction and implement emergency protective measures, we could prevent the attack and rescue a substantial amount of funds.

        \section{Discussion} \label{sec:disc}

\subsection{Vulnerability and Invariant} \label{sec:bug_invariant}

This section discusses the scope of smart contract vulnerabilities detected by SmartOracle and their violated invariants.

\noindent\textbf{Reentrancy}~\cite{bose2022sailfish} is a common smart contract vulnerability that allows attackers to re-enter the contract for stealing funds during external calls. However, re-entering the contract may violate the invariants between \textit{Pre-SubCall} and \textit{Post-SubCall} of the external call.

\noindent\textbf{Pribilege Escalation}~\cite{ghaleb2023achecker} means the contract lacks a complete access control check. Since many contracts use the \textit{whitelist} to record privileged users, triggering this bug may violate the invariant \texttt{\small msg.sender} $\in$ \texttt{\small whiteList}.

\noindent\textbf{Price Manipulation}~\cite{kong2023defitainter} is always caused by the poor implementation of the price oracle, which allows the attacker to manipulate the token price to make a profit. Exploiting this bug would abnormally change the token balances and violate their related invariants.

\noindent\textbf{Business Logic Flaw} strongly relies on the specific contract functionalities. SmartOracle can mine the essential invariants that fit the complicated contract functionalities, which serve as application-specific oracles for vulnerability detection. As in the example shown in Section~\ref{sec:motivation_challenges}, SmartOracle mines the essential invariant between the function parameter and the change of token balance from benign transactions, which are violated by the attack transactions triggering the vulnerability.

\subsection{Threats to Validity}

\noindent\textbf{External Validity}. The invariant mining procedure relies on the historical transactions of smart contracts, which prevents its application to the contract without any transactions. However, in our dataset, we find that about 80\% ($\frac{52}{65}$) contracts had over 100 transactions before being attacked, and 49\% ($\frac{32}{65}$) contracts had more than 1000 transactions, indicating that there are adequate historical transactions for invariant mining. Furthermore, the unit tests designed by developers can also provide transactions for invariant mining. Both indicate that our approach is available in most scenarios. 

\noindent\textbf{Internal Validity}. Manually collecting real-world incidents, locating their vulnerable contracts, and related attack transactions require significant labor, which may involve some mistakes. To ensure accuracy, all manual processes were performed by at least two experienced researchers. Additionally, although the dataset only contains 65 vulnerable contracts, they are representative and diverse from real-world DApps, so this dataset can evaluate the practical effectiveness of tools.
	\section{Related Work} \label{sec:related_work}

\subsection{Vulnerability Detection of Smart Contracts}

Numerous approaches have been proposed to detect smart contract vulnerabilities. Some approaches use general oracles to detect vulnerabilities~\cite{luu2016making, su2022effectively, feist2019slither, kong2023defitainter, zhou2020ever}. Luu et al.~\cite{luu2016making} originally proposed a symbolic executor for smart contracts, namely Oyente, that detects vulnerabilities with low-level and general oracles. Zhou et al.~\cite{zhou2020ever} manually extract patterns from known adversarial transactions, based on which they reveal the adversarial transactions by matching the logs with the patterns. They demonstrate great performance in revealing implementation vulnerability (e.g., integer overflow).

Meanwhile, some approaches use user-specific oracles to verify smart contracts~\cite{permenev2020verx, duan2022towards}. For example, Duan et al.~\cite{duan2022towards} proposed \textsc{VetSC} that utilizes model checking to vet smart contracts based on user-provided specifications. They can effectively detect vulnerabilities related to contract functionalities but rely on manual effort. In addition, Liu et al.~\cite{liu2022finding} proposed SPCon, which mines the roles of addresses from transactions and detects permission bugs in smart contracts. Similarly, SmartOracle also extracts rules from transactions, but for more kinds of vulnerabilities other than permission bugs.



\noindent\textbf{Difference.} Most existing tools use general or user-specified oracles to deal with vulnerabilities, which face ineffective detection or require too much manual effort. On the contrary, SmartOracle automatically extracts \textit{likely} invariants as application-specific oracles, which are suitable for detecting various kinds of vulnerability with complicated functionalities.

\subsection{Invariant Detection}

Due to its efficiency and scalability, invariant detection has been widely used in software testing and software verification~\cite{sahoo2013using, he2020structure, fioraldi2021use}. 
Among them, Daikon~\cite{ernst2007daikon} is a commonly used basic approach that dynamically extracts \textit{likely} invariants from running programs with a pattern-based inference engine. Based on Daikon, various approaches related to invariant detection are proposed~\cite{lemieux2015general, schuler2009efficient, beschastnikh2011leveraging, perkins2004efficient}. For example, Perkins et al.~\cite{perkins2004efficient} proposed an incremental algorithm to improve the efficiency and scalability of dynamic invariant detection.
Liu et al.~\cite{liu2022invcon} proposed a dynamic invariant detector for Ethereum smart contracts based on Daikon, namely InvCon. However, InvCon is only suitable to extract the \textit{likely} invariants within ERC20 token contracts, which limits its usage. 
Wang et al. proposed SmartInv~\cite{wang2024smartinv}, which uses a Large Language Model (LLM) to construct contract invariants based on source code and code comments. SmartInv requires labeled contracts to fine-tune the model. However, code comments may not be reliable or inconsistent with the code, which would interfere with the analysis.

\noindent\textbf{Difference.} Compared to existing approaches, SmartOracle is a more suitable and practical invariant detector for smart contracts. SmartOracle extracts more fine-grained invariants, uses a threshold mechanism to improve usefulness, and provides automated run-time verification. Thus, SmartOracle has more ability to handle various smart contract vulnerabilities.
	\section{Conclusion} \label{sec:conclusion}

In this paper, we propose an automatic invariant detector, namely SmartOracle, that generates fine-grained invariants as application-specific oracles for detecting smart contract vulnerabilities. Specifically, SmartOracle first parses the smart contract and its transactions to extract execution traces. Then, SmartOracle utilizes a threshold-based algorithm to mine \textit{likely} invariants, based on which it detects smart contract bugs by checking whether the transactions violate the invariants. In experiments, SmartOracle outperforms existing dynamic invariant detection, detects 50\% more ERC20 invariants and achieves 96\% precision. Furthermore, SmartOracle shows its effectiveness in generating application-specific oracles for detecting real-world vulnerabilities. It detects 31 vulnerable contracts and their related 466 transactions that violated mined invariants. 


	\maketitle\maketitle
	\balance
	\bibliographystyle{IEEEtran}

	
\end{document}